\RequirePackage{fix-cm}
\documentclass{svjour3}

\smartqed
\usepackage{amsmath}
\usepackage{mathptmx}

\usepackage[utf8]{inputenc}
\usepackage{graphicx}
\usepackage{color}
\usepackage{datetime}

\usepackage{ifpdf}

\newtheorem{observation}{Observation}

\def\R{\mathbf{R}}
\def\Z{\mathbf{Z}}
\def\NP{{\ensuremath{\text{\textsc{NP}}}}}

\DeclareMathOperator{\dem}{dem}

\begin{document}

\title{A Polynomial Time Approximation Algorithm\\ for the \\
  Two-Commodity Splittable Flow Problem
\thanks{The first author was supported by a PAKT project of the German
  Leibniz Society.}}
\titlerunning{Two-Commodity Splittable Flow}
\author{Elke Eisenschmidt \and Utz-Uwe Haus}
\institute{%
  Elke Eisenschmidt \at
  Institut für Mathematische Optimierung,
  Otto-von-Guericke Universität Magdeburg, 
  Universitätsplatz 2
  39106 Magdeburg, Germany\\
  \email{eisensch@mail.math.uni-magdeburg.de}
  \and
  Utz-Uwe Haus \at
  Institut für Operations Research, ETH Zürich,
  Rämistrasse 101 8092 Zürich, Switzerland\\
  Tel.: +41-44-63-39355, 
  Fax: +41-44-63-21025\\
 \email{uhaus@math.ethz.ch} 
}
\date{Received: date / Accepted: date}
\maketitle

\begin{abstract}
  We consider a generalization of the unsplittable maximum
  two-commodity flow problem on undirected graphs where each commodity
  $i\in\{1,2\}$ can be split into a bounded number $k_i$ of equally-sized chunks
  that can be routed on different paths. We show that in contrast to
  the single-commodity case this problem is \NP-hard, and hard to
  approximate to within a factor of $\alpha>1/2$. We
  present a polynomial time $1/2$-approximation algorithm for the case
  of uniform chunk size over both commodities and show that for even
  $k_i$ and a mild cut condition it can be modified to yield an exact
  method. The uniform case can be used to derive a $1/4$-approximation
  for the maximum concurrent $(k_1,k_2)$-splittable flow without chunk
  size restrictions for fixed demand ratios.

\end{abstract}

\section{Introduction}
\label{sec:introduction}

We consider a generalization of the unsplittable maximum two-commodity
flow problem~\cite{kleinberg:96} on an undirected capacitated graph
$G=(V,E)$ introduced by~\cite{baier-koehler-skutella:02} where each
commodity $i$ can be split into a bounded number $k_i$ of chunks (of
potentially different size) which can be routed on different paths
($k$-splittable flow problem). This problem is \NP-hard even for one
commodity and $k=2$, see~\cite{baier-koehler-skutella:05}, unless
extra restrictions are imposed.

In the following we will always work with an undirected graph
$G=(V,E)$, with $s_1,s_2\in V$ the sources, and $t_1,t_2\in V$ the
sinks of two commodities of flow.

\begin{definition}[splittable flow]
  Let $G=(V,E)$ be an undirected graph with edge capacities $u_e$
  $(e\in E)$, and let $s_1,s_2\in V$ be the
  sources and $t_1,t_2\in V$  be the sinks of two commodities of
  flow, and $k_1,k_2$ two nonnegative integers. A
  \emph{$(k_1,k_2)$-splittable flow} is a two-commodity flow
  respecting the edge capacities using
  $k_1$ $s_1$--$t_1$-paths for commodity 1 and $k_2$
  $s_2$--$t_2$-paths for commodity 2.
\end{definition}

Since we allow that a path can be used multiple times and flow on
certain paths can  be equal to $0$, the notion of
$k_1,k_2$-splittability includes the case where `at most $k_i$ paths'
may be used for commodity $i$.
 
However, in many applications commodities cannot be split into arbitrarily
sized chunks, which  puts restrictions on the allowable
flow values of the splittable flow. One reasonable restriction is to
require that for each commodity the individual flows need to have the
same flow value. The paths in a splittable flow do not
need to be different, therefore integral multiples of such `chunk-sized'
transport can be accomodated on the same path.

\begin{definition}[bi-uniform splittable flow]
  A $k_1,k_2$-splittable flow is called \emph{bi-uniform} if the flow
  values of the paths for each commodity are the same.
\end{definition}

Note that with uniformity restrictions, a $0$-flow on some path will
force all flows for the respective commodity to be $0$. Thus the
problem reduces to a problem with one commodity less.

In the single-commodity case Baier et
al.~\cite{baier-koehler-skutella:05} show that assuming uniformity
makes the problem solvable in polynomial time. We will show that this
is not the case for two commodities, not even if we ask for uniformity across
both commodities. The latter restriction is also not artificial: Imagine
that each commodity models a different service level, but the
underlying good is divisible only in the same fashion, e.g., into
packet size or base channel bandwidth in a telecommunication network.

\begin{definition}[totally uniform splittable flow]
  A $k_1,k_2$-splittable flow is called \emph{totally uniform} if the flow
  values of all paths for all commodities are the same.
\end{definition}

\begin{lemma}
  The following problems are \NP-hard:
  \begin{itemize}
  \item Maximize the flow per path of a totally uniform 
    $k_1,k_2$-splittable flow,
  \item Maximize the sum $x+y$ where $x$ ($y$) is the flow per path of
    commodity $1$ (of commodity $2$) of a bi-uniform
    $k_1,k_2$-splittable flow,
  \item  Maximize the total non-uniform $k_1,k_2$-splittable
    flow.
  \end{itemize}
\end{lemma}

\begin{proof}
  The variant without any uniformity constraints was shown to be \NP-hard
  by Baier et~al.~\cite{baier-koehler-skutella:05}, as noted above.

  We will show that the integral 2-commodity flow problem with unit
  capacities is reducible to both the totally uniform and the
  bi-uniform $k_1,k_2$-splittable flow problem.
  
  Let $G=(V,E)$ with sources $s_1,s_2$ and sinks $t_1,t_2$, identical
  capacities of $1$ on each edge $e\in E$, and demands $d_1,d_2\in\Z_{\geq0}$
  be given. Evan et al.~\cite{even-itai-shamir:76} show that asking
  whether there exists an integral 2-commodity flow satisfying the
  demands for such a graph is \NP-hard (even
  though  the capacities are all $1$). 

  Let $k_1=d_1$ and $k_2=d_2$. Solving the totally uniform
  (respectively, the bi-uniform) $k_1,k_2$-splittable flow problem on
  $G$ yields a solution composed of $k_1$ paths for commodity $1$ and
  $k_2$ paths of commodity $2$. All paths have the same flow value $x$
  (resp.: $x$ and $y$) for the commodities. If $x=1$ (resp.: $x+y=2$)
  then we have found an integral two-commodity flow satisfying the
  demands.  If $x<1$ (resp.: $x+y<2$) then there exists no
  integral two-commodity flow satisfying the demands: Assume there
  were an integral two-commodity flow satisfying $d_1$ and $d_2$, then
  without loss of generality we can assume that it exactly satisfies
  the demands. Then it is, however, also a $k_1,k_2$-splittable flow
  -- since each edge carries an integral flow, i.e. a value of $0$ or $1$,
  we can split it into exactly $k_1$ and $k_2$ paths for commodity $1$
  and $2$, respecitively.  In particular, the flow value of each of
  the paths is $1$, contradicting $x<1$ (resp.: $x+y<2$).
\qed\end{proof}

\begin{figure}
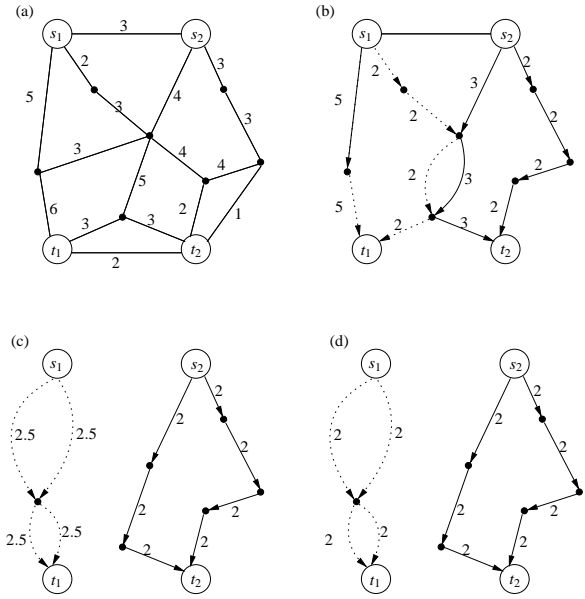

  \centering
  \resizebox{!}{.4\textheight}{\ifpdf\input{introductory-example.pdftex_t}\else\input{introductory-example.pstex_t}\fi}

  \caption{Variants of splittable flows: a) The graph, b) a maximal unconstrained $(2,2)$-splittable flow, c) a maximal bi-uniform $(2,2)$-splittable flow, d) a maximal totally uniform $(2,2)$-splittable flow.}
  \label{fig:intro}
\end{figure}
Re-reading the proof we can see that the flow value $x$ (resp.: $x$
and $y$) on the paths
of an optimal $k_1,k_2$-splittable totally uniform (resp.: bi-uniform) flow
solution on the class of instances considered can never lie in the
open interval $(1/2,1)$, since such a flow can always be increased to
$1$. A flow value of $1/2$ could be possible, if some edge is used by
two paths (this corresponds to fractional, and therefore
half-integral, solvability of the 2-commodity integral flow
problem~\cite{hu:63}). Hence, any $\alpha$-approximation algorithm of
the totally uniform $k_1,k_2$-splittable flow problem with
$\alpha>1/2$ will also answer solve the integral 2-commodity flow
problem: approximate solutions with flow $x>1/2$ must correspond to
``YES''-instances of the 2-commodity integral multicommodity flow
problem, and approximate solutions with flow $x\leq 1/2$ to
``NO''-instances.  This yields the following:

\begin{corollary}\label{lem:0.5-hardness}
  It is \NP-hard to approximate the maximum totally uniform
  $k_1,k_2$-splittable flow problem  to within a
  factor of $\alpha>1/2$, even for graphs with unit capacities.
\end{corollary}

\begin{figure}
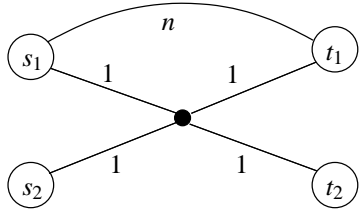

  \centering
  \ifpdf\input{uniformity-example.pdftex_t}\else\input{uniformity-example.pstex_t}\fi
  \caption{A graph with maximal $1,1$-splittable flows of different
    values depending on the version of uniformity: The  maximal totally uniform $1,1$-splittable flow has a value
$x+x = 2$; the maximal bi-uniform $1,1$-splittable flow has a value 
$x+y = n+1$.}
  \label{fig:uniformity-example}
\end{figure}
It would be tempting to try and use totally uniform splittable flows
to approximate bi-uniform splittable flows, but, as
Figure~\ref{fig:uniformity-example} shows, this is not possible.

\medskip

There are various notions of maximality for splittable flows that
in general do not yield the same solutions.

\begin{definition}[maximality notions]
  Let $(f^1_1,\dots,f^1_{k_1},f^2_{1},\dots,f^2_{k_2})$ be a
  $k_1,k_2$-splittable two-commodity flow in a graph $G$. It is called 
  \begin{itemize}
  \item \emph{maximal total flow} if it is optimal for 
    $$\max
    \sum_{i=1}^{k_1} f^1_i+\sum_{i=1}^{k_2}f^2_i,$$

  \item \emph{maximal concurrent flow} if for some given demand 
    parameters $d_1,d_2\in\R_{\geq 0}$ it is optimal for
    $$\max_{\text{$f$ a $k_1,k_2$-splittable 2-c-f}\quad}
    \min_{i\in\{1,2\}}
    \tfrac{1}{d_i}\sum_{j=1}^{k_i}f_j^i,
    $$
  \item \emph{maximal flow} if it is 
    optimal for
    $$ \sum_{i=1}^2\max_{j\in\{1,\dots,k_i\}} f_i^j$$
  \end{itemize}
  among all feasible $k_1,k_2$-splittable two-commodity flows of $G$.
\end{definition}

We will mostly be concerned with maximal totally uniform or bi-uniform
flows, except for Section~\ref{sec:appr-nonuniform}, where we study
maximal concurrent flow. In the former case the objective function
simplifies to $\max x+y$ where $x$ and $y$ are the flow
values per path for the two commodities (and $x=y$ for totally uniform
flows).

\section{Bi-uniform and totally uniform splittable flows}
\label{sec:totally-uniform}

From classical multicommodity flow theory we know that the maximum
multicommodity flow is bounded by the minimum multicommodity cut. In
the single-commodity case this bound is tight, as asserted by the
max-flow min-cut theorem. In \cite{baier-koehler-skutella:05} this was
extended to the case of single-commodity uniform $k$-splittable
$s$--$t$-flows:
\begin{definition}[minimum k-cut]
  Let $S\subseteq V$ with $s\in S$ and $t\in V\setminus S$ be a cut in
  $G=(V,E)$, and define 
  \begin{equation}
    \label{eq:ck}
    c_k(S):=\max\{x\in\R_{\geq 0}\;:\; 
    \sum_{e\in\delta(S)}n(e)=k, n(e)\in\Z_{\geq0}\text{ and }n(e)x\leq u_e \text{ for all }e\in\delta(S)\}
  \end{equation}
  as the maximum item size such that $k$ elements of equal size fractionally 
  fit into the bins created by the edge capacities of $\delta(S) := \{(u,v) \in E \colon (u \in S \land v \notin S) \text{ or } (v \in S \land u \notin S)\}$.
  Then
  \begin{equation}
    \label{eq:ckG}
    c_k(G)=\min\{c_k(S)\;:\; S\subseteq V, s\in S, t\in V\setminus S\}
  \end{equation}
  is called \emph{minimum $k$-cut value} of $G$.  
\end{definition}

Baier et al.~\cite{baier-koehler-skutella:05} show that the value of the maximum
uniform $k$-splittable $s$--$t$-flow in $G$ equals the minimum $k$-cut
value $c_k(G)$.

One can consider a similar approach for the two-commodity flow
problem, i.e. consider a similar packing problem for two different
items:
\begin{equation}
  \label{eq:2-obj-packing}
  \begin{array}{rlll}
    \max& x+y\\
    \text{s.t.}&n_1(e)x+n_2(e)y&\leq u_e&\forall e\in\delta(S)\\
               &\sum_{e\in\delta(S)}n_1(e)& \geq k_1&\text{if }(s_1\in
               S,t_1\in V\setminus S)\text{ or }
               (t_1\in S,s_1\in V\setminus S) \\
               &\sum_{e\in\delta(S)}n_2(e)& \geq k_2&\text{if }(s_2\in
               S,t_2\in V\setminus S)\text{ or }
               (t_2\in S,s_2\in V\setminus S) \\
               &n_1(e),n_2(e)\in\Z_{\geq0}&&\forall e\in\delta(S)\\
               &x,y\in\R_{\geq0}               
  \end{array}
\end{equation}

\begin{proposition}[cut bound]\label{prop:ck1k2-bound}
  For a graph $G=(V,E)$ and each cut $S\subseteq V$ with
  $s_1,s_2\in S$ and $t_1,t_2\in V\setminus S$ the
  two-commodity bin packing problem~(\ref{eq:2-obj-packing}) provides
  an upper bound for the value of a bi-uniform $k_1,k_2$-splittable
  flow on $G$, but this minimum cut bound need not be tight.
\end{proposition}

\begin{proof}
  Clearly, the flow values $(x^*,y^*)$ of a valid bi-flow which is
  split according to $n_1^*,n_2^*$ have to satisfy the conditions
  of~(\ref{eq:2-obj-packing}), hence the optimum
  of~(\ref{eq:2-obj-packing}) provides an upper bound.

  \begin{figure}
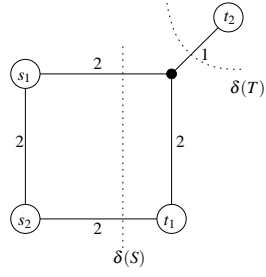

    \centering
    \resizebox{!}{.18\textheight}{\ifpdf\input{2cut-counterexample.pdftex_t}\else\input{2cut-counterexample.pstex_t}\fi}
    \caption{A two-commodity digraph with maximum $1,1$-splittable
      flow of value $2$ but best 1-cut packing bound of $4$ (realized
      by $S$) and best 2-cut packing bound of $3$ (realized by $S$ and
      $T$).}
    \label{fig:packing-bound-gap}
  \end{figure}
  The graph in Figure~\ref{fig:packing-bound-gap} for $k_1=1$ and
  $k_2=1$ allows a maximal bi-uniform flow of value $x+y=2$, but
  minimizing~(\ref{eq:2-obj-packing}) over all cuts only yields a
  bound of $4$.
\qed \end{proof}

One might consider adding two independent sets of cut constraints to
the system~\eqref{eq:2-obj-packing}, in an attempt to allow one cut to
bound $x$ well, and the other to bound $y$ well, and thus obtain a
stronger cut bound. Clearly, such a formulation will not be weaker
than~\eqref{eq:2-obj-packing}, but it still does not yield a tight cut
bound in general, as we also illustrate in
Figure~\ref{fig:packing-bound-gap}: All possible cuts have values of
either $1$, $4$, $5$ or more. The cuts $S$ and $T$ in the Figure are
therefore exemplary best cuts, and yield only a bound of $1$ for $y$
(cut $T$), and $2$ for $x$ (cut $S$), giving a joint bound of $x+y\leq
3$. We therefore only consider system~\eqref{eq:2-obj-packing} with
one cut.

Note that \eqref{eq:2-obj-packing} is a mixed-integer nonlinear optimization
program which we cannot expect to directly use for solving the problem.
If, however, one assumes uniformity across commodities, the
bin-packing problem~(\ref{eq:2-obj-packing}) turns out to be useful
even in the two-commodity case. Let $\{s_1,s_2,t_1,t_2\}$ be the
sources and destinations of the $k_1,k_2$-splittable totally uniform
two-commodity flow problem and consider a set of nodes $S\subseteq V$.
We define 
\begin{equation}
  \label{eq:dem}
  \dem(S)= \begin{cases}
    k_1     & (s_1\in S \land \{s_2,t_1,t_2\}\not\subseteq S)\text{ or }
    (t_1\in S \land \{s_1,s_2,t_2\}\not\subseteq S)\\
    k_2     & (s_2\in S \land \{s_1,t_1,t_2\}\not\subseteq S)\text{ or }
    (t_2\in S \land \{s_1,s_2,t_1\}\not\subseteq S)\\
    k_1+k_2 & (s_1,s_2\in S \land \{t_1,t_2\}\not\subseteq S)\text{ or }
    (t_1,t_2\in S \land \{s_1,s_2\}\not\subseteq S)\\
    k_1+k_2 & (s_1,t_2\in S \land \{s_2,t_1\}\not\subseteq S)\text{ or }
    (s_2,t_1\in S \land \{s_1,t_2\}\not\subseteq S)\\
    0       & \text{otherwise}
  \end{cases}
\end{equation}
the demand necessarily crossing  $\delta(S)$ in a feasible flow.

Then~(\ref{eq:2-obj-packing}) can be rewritten as
\begin{equation}
  \label{eq:2-obj-tu-packing}
  \begin{array}{rlll}
    c_{k_1,k_2}(S):=\max& x\\
    \text{s.t.}&n(e)x&\leq u_e&\forall e\in\delta(S)\\
               &\sum_{e\in\delta(S)}n(e)& \geq \dem(S)&\\
               &n(e)\in\Z_{\geq0}&&\forall e\in\delta(S)\\
               &x\in\R_{\geq0}               
  \end{array}
\end{equation}

We denote by $c_{k_1,k_2}(G)$ the minimum such cut value:
\begin{equation}
  \label{eq:ck1k2-G}
  c_{k_1,k_2}(G):=\min_{S\subseteq V,\dem(S)\neq 0}c_{k_1,k_2}(S)  
\end{equation}

\begin{figure}
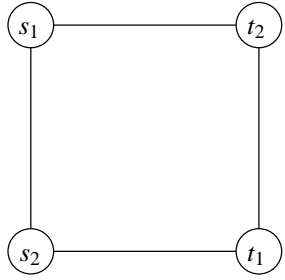

  \centering
  \ifpdf\input{critical-minor.pdftex_t}\else\input{critical-minor.pstex_t}\fi
  \caption{Forbidden minor for integrality of two-commodity flow problems.}
  \label{fig:bad-minor}
\end{figure}
\begin{lemma}\label{lem:2k-totally-uniform}
  Let $G=(V,E)$ be an undirected graph with edge capacities
  $u\in\Z_{\geq 0}^{|E|}$ and let $k_1,k_2\in\Z_{\geq 0}\setminus
  \{0\}$. Then there exists a $2k_1,2k_2$-splittable totally uniform
  flow with value $(k_1+k_2)c_{k_1,k_2}(G)$. Furthermore, if the graph
  in Figure~\ref{fig:bad-minor} is not a minor of $G$, there exists a
  $k_1,k_2$-splittable totally uniform flow with this value.
\end{lemma}
\begin{proof}
  Let $x=c_{k_1,k_2}(G)$ be the minimum $k_1,k_2$-cut value as defined
  in~(\ref{eq:ck1k2-G}) and let $n \in \Z_{\geq 0}^{|E|}$ be the 
  corresponding feasible solution.  We construct an auxiliary graph $G'=(V,E)$
  with edge capacities $u_e'=\lfloor\tfrac{u_e}{x}\rfloor$.

  Now
  consider the two-commodity flow problem on $G'$ with demands
  $d_1=k_1$ and $d_2=k_2$. As $n(e)x \leq u_e$ for all $e \in E$,
  we have $n(e) \leq \frac{u_e}{x}$. As $n(e) \in \Z_{\geq 0}$ we
  can round the right-hand side of this inequality. Therefore,
  $n(e) \leq \lfloor\tfrac{u_e}{x}\rfloor = u_e'$. 
  
  In particular for every $S\subseteq V$, $\sum_{e\in\delta(S)} u_e'
  \geq \sum_{e \in \delta(S)}{n(e)} \geq \dem(S)$.  According to Hu's
  two-commodity flow theorem~\cite{hu:63}, there exists a
  half-integral solution for demands $d_1 = k_1$, $d_2=k_2$.  This
  half-integral solution can be constructed in polynomial time, see
  e.g.~\cite[Theorem 71.1b]{Schrijver:03:CoOptC}. Regular
  flow-decomposition techniques yield a solution with $2k_1$ paths for
  commodity 1 and $2k_2$ paths for commodity 2, each carrying a flow
  of $1/2$.
  
  On the original graph $G$ we assign these paths a flow of
  $\tfrac{1}{2}x$. We thus obtain a feasible two-commodity flow on $G$
  with total flow of $(k_1+k_2)x$. 

  If the graph in Figure~\ref{fig:bad-minor} is not a minor of $G$ 
  there even exists an integral two-commodity flow solution instead of a
  half-integral one (see e.g.~\cite[Theorem 71.2]{Schrijver:03:CoOptC}),
  which directly yields a $k_1,k_2$-splittable solution with the 
  same value.
\qed \end{proof}

The factor of $2$ for the number of paths in
Lemma~\ref{lem:2k-totally-uniform} is sometimes best possible, as the
following example shows.
\begin{example}
  Consider the graph in Figure~\ref{fig:bad-minor} with edge
  capacities $u_e=1$ for all edges, and $k_1=1=k_2$. Then clearly
  $c_{1,1}(G)=1$, but there is no $1,1$-splittable totally uniform
  flow with a value of $(k_1+k_2)c_{k_1,k_2}=2\cdot 1=2$. However,
  there exists a $2,2$-splittable totally uniform flow with the value
  $(2+2)\cdot(1/2)=2$.
\end{example}

For even $k_1$ and $k_2$ dividing these parameters by $2$ and applying
Lemma~\ref{lem:2k-totally-uniform} obviously always yields a feasible
solution of the $k_1,k_2$-splittable totally uniform flow problem.
One could hope that it would be possible to use
Lemma~\ref{lem:2k-totally-uniform} for $\bar{k_1}=k_1/2$ and
$\bar{k_2}=k_2/2$ when $k_1$ and $k_2$ are even to compute a
maximum $k_1,k_2$-splittable flow. The next example shows, however,
that this is not possible in general.

\begin{example}
 Let $k_1=k_2=2$ and consider the graph in
  Figure~\ref{fig:factor-2}.
  \begin{figure}
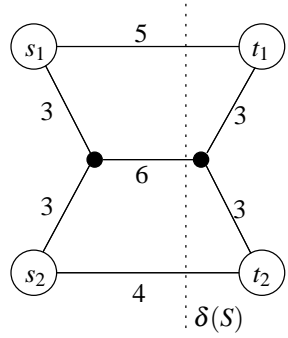

    \begin{center}
    \ifpdf\input{example-2.pdftex_t}\else\input{example-2.pstex_t}\fi
    \end{center}
    \caption{A graph with optimal $2,2$-splittable totally uniform
      flow of value $12$, and optimal $1,1$-splittable totally uniform
      flow of value $8$.}
    \label{fig:factor-2}
  \end{figure}
  Here $c_{1,1}(G)=c_{1,1}(S)=4$ and the corresponding
  auxiliary graph has precisely one integral solution. However,
  $c_{2,2}(G)=c_{2,2}(S)=3$, and there is indeed a $2,2$-splittable totally
  uniform solution yielding a total flow of $12$.
\end{example}

It is easy to obtain the necessary condition for the flow obtained by
Lemma~\ref{lem:2k-totally-uniform} to be maximal though. We start with
the following observation.

\begin{observation}
  For a graph $G=(V,E)$ with edge capacities $u_e\in\Z_{\geq 0}$ for
  all $e\in E$ and nonnegative integers $k_1,k_2$ it holds that
  \begin{equation}
    \label{eq:ck-ck/2}
    2 c_{2k_1,2k_2}(G)\geq c_{k_1,k_2}(G).    
  \end{equation}
\end{observation}

This follows from the fact that a feasible flow $x$ for $c_{k_1,k_2}$
in~\eqref{eq:2-obj-tu-packing} always yields a feasible flow $x/2$
for $c_{2k_1,2k_2}$ in~\eqref{eq:2-obj-tu-packing}. Hence $2
c_{2k_1,2k_2}(G)$ can not be smaller than $c_{k_1,k_2}$. Tightness
in~\eqref{eq:ck-ck/2} is the necessary condition for
applicability of the following Lemma:

\begin{lemma}
  Let $k_1,k_2\in2\Z_{\geq0}$ be even integers, $G=(V,E)$ with edge capacities
  $u_e\in\Z_{\geq0}$ and assume
  $2c_{k_1,k_2}(G)=c_{k_1/2,k_2/2}(G)$. Then an optimal solution of the
  $k_1,k_2$-splittable totally uniform flow problem can be obtained by
  applying Lemma~\ref{lem:2k-totally-uniform} to $G$, $k_1/2$, and $k_2/2$.
\end{lemma}
\begin{proof}
  Using Lemma~\ref{lem:2k-totally-uniform} for $k_1/2$ and $k_2/2$
  yields a $k_1,k_2$-splittable totally uniform flow where each path
  carries a flow of $1/2 c_{k_1/2,k_2/2}(G)$. Since $2
  c_{k_1,k_2}(G)=c_{k_1/2,k_2/2}(G)$  by assumption and
  $c_{k_1,k_2}(G)$ is an upper bound by
  Proposition~\ref{prop:ck1k2-bound}, the claim follows.
\qed \end{proof}

We will now show that the value of $c_{k_1,k_2}(G)$ can be computed in
polynomial time, allowing us to check whether~(\ref{eq:ck-ck/2}) is
satisfied. Furthermore, knowing the value of $c_{k_1,k_2}(G)$ allows
us to compute a factor $1/2$-approximation for the maximum totally
uniform flow problem in the general case.

\begin{lemma}
  The value $c_{k_1,k_2}(G)$ can be computed in polynomial time
  $\mathcal{O}((k_1+k_2)|E|\log|E|)$.
\end{lemma}

\begin{proof}
  To compute $c_{k_1,k_2}(G)$ we have to find the minimum of
  $c_{k_1,k_2}(S)$ over all cuts $S$ in $G$ with $\dem(S)\neq 0$.

  We can distinguish four cases according to~(\ref{eq:dem}), depending
  on which subset of $\{s_1,s_2,t_1,t_2\}$ is contained in $S$,
  yielding four relevant values of $\dem(S)$:
  \begin{enumerate}
  \item $\dem(S)=k_1$. Then $c_{k_1,k_2}(S)=c_{k_1}(S)$.
  \item $\dem(S)=k_2$. Then $c_{k_1,k_2}(S)=c_{k_2}(S)$.
  \item $\dem(S)=k_1+k_2$ because $s_1,s_2\in S$ and $t_1,t_2\in
    V\setminus S$ (or symmetrically $t_1,t_2\in S$ and $s_1,s_2\in
    V\setminus S$). Then $c_{k_1,k_2}(S)=c_{k_1+k_2}(S)$.
  \item $\dem(S)=k_1+k_2$ because $s_1,t_2\in S$ and $s_2,t_1\in
    V\setminus S$ (or symmetrically $s_2,t_1\in S$ and $s_1,t_2\in
    V\setminus S$). Then $c_{k_1,k_2}(S)=c_{k_1+k_2}(S)$.
  \end{enumerate}
  Determining the value $c_{k_1,k_2}(G)$ thus amounts to determining
  the minimum of three single-commodity $l$-cut values
  (for $l\in\{k_1,k_2,k_1+k_2\}$) w.r.t. certain auxiliary
  graphs. The auxiliary graphs are presented in
  Figure~\ref{fig:aux-graphs}. In each case we have to determine a
  $l$-cut value for a $s'$--$t'$-flow.
  \begin{figure}
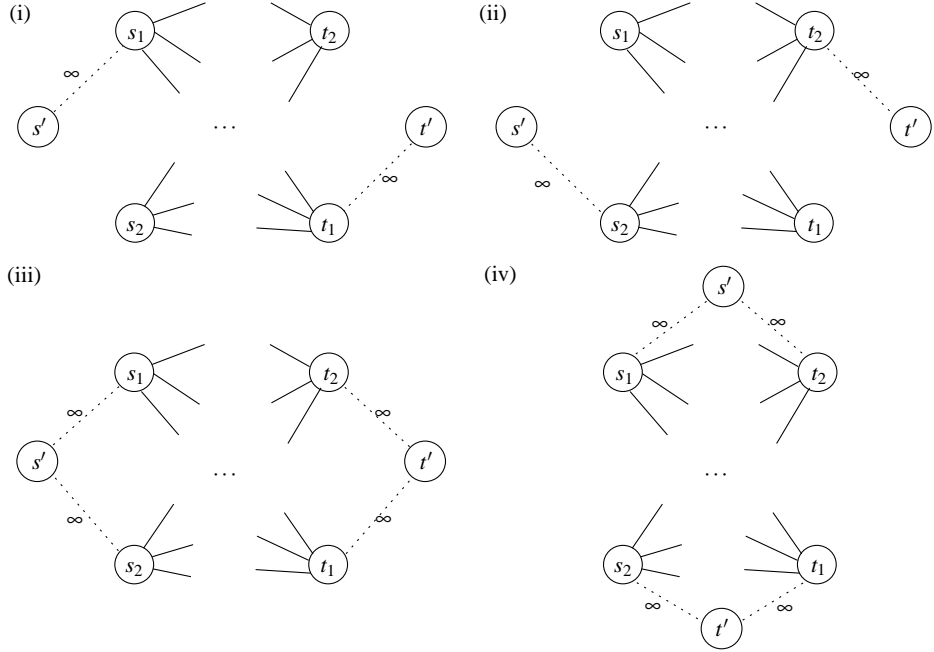

    \centering
    \resizebox{\textwidth}{!}{\ifpdf\input{aux-graphs.pdftex_t}\else\input{aux-graphs.pstex_t}\fi}
    \caption{The auxiliary graphs for determining a minimum
      $k_1,k_2$-cut on G. The auxiliary edges are displayed as dotted
      lines and have capacity $\infty$.}
    \label{fig:aux-graphs}
  \end{figure}

  Computing an individual value $c_k$ can be done in time
  $O(k|E|\log|E|)$ using the algorithm of~\cite{baier-koehler-skutella:05}.

\qed \end{proof}

So far we have shown that in the special case where the graph in
Figure~\ref{fig:bad-minor} is not a minor of $G$ and equality holds
in~(\ref{eq:ck-ck/2}) we can solve the maximum totally
uniform flow problem exactly using two calls to a single-commodity
integral flow algorithm.

In the general case a factor $1/2$ approximation
is achievable in polynomial time. Given
Corollary~\ref{lem:0.5-hardness} this is best possible unless $\mathcal{P}=\NP$.

\begin{theorem}\label{thm:approx-biuniform}
  Consider the
  $k_1,k_2$-splittable totally uniform 2-commodity flow problem on an
  undirected graph $G=(V,E)$  with edge capacities
  $u_e\in\Z_{\geq0}$ for $e\in E$. Then a $1/2$-approximation for the
  maximal totally uniform flow can be computed in polynomial time.
\end{theorem}

\begin{proof}
  This is a direct consequence of
  Lemma~\ref{lem:2k-totally-uniform}: It yields a feasible
  two-commodity flow composed of $2k_1$ and $2k_2$ paths with total
  flow value of $(k_1+k_2)c_{k_1,k_2}$. Dropping $k_1$ paths carrying
  commodity $1$ and dropping $k_2$ paths carrying commodity $2$ we
  obtain a $k_1,k_2$-splittable solution with totally uniform
  path-flow across commodities and a total flow of
  $\tfrac{1}{2}(k_1+k_2)c_{k_1,k_2}$. This is at least a $1/2$
  approximation since $c_{k_1,k_2}$ is an upper bound on the path flow.

\qed \end{proof}

\section{Approximating nonuniform concurrent flow}
\label{sec:appr-nonuniform}

Finally we will show that a general $k_1,k_2$-splittable two-commodity flow
can be approximated with the help of uniform flows.

\begin{theorem}\label{thm:approx-nonuniform}
  Let $G=(V,E)$ be an undirected graph with edge capacities
  $u_e\in\Z_{\geq0}$ for all $e\in E$. Let $k_1,k_2\in\Z_{\geq0}$ be
  integral parameters. A maximal  totally uniform $k_1,k_2$-splittable
  flow provides a $\tfrac{1}{2}$-approximation of a maximal concurrent
  $k_1,k_2$-splittable flow for a demand ratio $d_1/d_2=k_1/k_2$.
\end{theorem}

\begin{proof}
  Theorem~13 in~\cite{baier-koehler-skutella:05} states that every
  maximal bi-uniform $k_1,k_2$-splittable flow is a
  $\frac{1}{2}$-approximation of a maximal $k_1,k_2$-splittable
  flow. We will show that for $d_1/d_2 = k_1/k_2$, a maximal
  bi-uniform flow is in fact totally uniform.

  Let ${\mathcal P}_i$ denote the set of $s_i$-$t_i$ paths of commodity 
  $i$ and consider the maximum concurrent bi-uniform 
  $k_1,k_2$-splittable flow problem for demands $d_1/d_2=k_1/k_2$:
 
\begin{equation}
\label{xyz}
\begin{aligned}
\max \quad & \lambda\\
\mbox{s.t.} \quad& \sum\limits_{\substack{p \in {\mathcal P}_1,\\ e \in p}}{x \, \delta_p} 
+ \sum\limits_{\substack{q \in {\mathcal P}_2,\\e \in q}}{y \, \delta_q} &\leq \;& u_e & \forall e \in E\\
& \sum\limits_{p \in {\mathcal P}_1}{\delta_p} & =\; & k_1\\
& \sum\limits_{q \in {\mathcal P}_2}{\delta_q} & =\; & k_2\\
& \lambda d_1 & =\; & k_1 \, x\\
&\lambda d_2 & =\; & k_2 \, y\\
& \delta_p, \delta_q \in \{0,1\} &&\;& \forall p \in {\mathcal P}_1, \, \forall q \in {\mathcal P}_2\\
& x, y, \lambda \in \R_{\geq0} \\
\end{aligned}
\end{equation}
The first set of inequalities ensures the edge capacities are
respected. The second and third set of equalities ensures that $k_1$
paths for commodity $1$ and $k_2$ paths for commodity $2$ are
used. The fourth and fifth set of inequalities finally relate the
demands $\lambda d_i$, of commodity $i$, to the flow for commodity
$i$, $k_1\,x$ and $k_2 \, y$, respectively. From these last two
equalities (and from $d_1/d_2=k_1/k_2$), we obtain that $x = y$ has 
to hold, and thus a feasible $k_1,k_2$-splittable bi-uniform flow is 
in fact totally uniform. 

Now we will show that a maximal totally uniform $k_1,k_2$-splittable
flow provides an optimal solution for the program \eqref{xyz}. Let
$x$ be the flow value on the $k_i$ paths of commdity $i$. Then 
$\bar{d}_1 := k_1x$ is the total flow of commodity $1$ and 
$\bar{d}_2 := k_2x$ is the total flow of commodity $2$. We will 
show that $\bar{d}_1 = \lambda d_1$ and $\bar{d}_2 = \lambda d_2$
for maximal $\lambda$.
We have $\bar{d}_1 = k_1x = \frac{d_1k_2}{d_2}x = \frac{\bar{d}_2}{d_2} d_1$
and thus $\bar{d}_2 = \frac{\bar{d}_1}{d_1}d_2$. Therefore, we
have to show that $\frac{\bar{d}_2}{d_2} = \frac{\bar{d}_1}{d_1}$
holds. But this follows directly from  
\begin{displaymath}
 \frac{\bar{d}_2}{d_2} =\frac{k_2x}{d_2} = \frac{k_1x}{d_1}= \frac{\bar{d}_1}{d_1}.
\end{displaymath}
Therefore, $\lambda =  \frac{k_1x}{d_1} =  \frac{k_2x}{d_2}$.
As $d_i$ and $k_i$ are fix, it is clear that a maximum value
of $x$ yields a maximal value of $\lambda$.

This concludes our proof: as a maximal $k_1,k_2$-splittable totally 
uniform flow is a maximal concurrent 
$k_1,k_2$-splittable bi-uniform flow for demand ratios $d_1/d_2 = k_1/k_2$, 
it provides a $\frac{1}{2}$ approximation for the maximal concurrent 
$k_1,k_2$-splittable flow for demand ratios $d_1/d_2 = k_1/k_2$.
\qed \end{proof}

As a direct consequence of applying both
Theorems~\ref{thm:approx-nonuniform} and~\ref{thm:approx-biuniform}
consecutively we obtain

\begin{corollary}
  Let $G=(V,E)$ be an undirected graph with edge capacities
  $u_e\in\Z_{\geq0}$ for all $e\in E$. Let $k_1,k_2\in\Z_{\geq0}$ be
  integral parameters. A $1/4$-approximation of a maximal 
  concurrent $k_1,k_2$-splittable flow can be computed in polynomial
  time for demand-ratios $d_1/d_2 = k_1/k_2$.
\end{corollary}
\bibliographystyle{spmpsci}
\bibliography{splittable-flow}

\end{document}